\documentclass[10pt]{article}
\usepackage[latin1]{inputenc}
\usepackage[T1]{fontenc}
\usepackage{amsmath}
\usepackage{amsfonts}
\usepackage{amssymb}
\usepackage{latexsym}
\usepackage[english]{babel}
\usepackage[dvips]{graphicx,color}
\usepackage{graphicx}
\usepackage{graphics}
\newcommand{\be}{\begin{equation}}
\newcommand{\ee}{\end{equation}}
\newcommand{\beq}{\begin{equation}}
\newcommand{\eeq}{\end{equation}}
\newcommand{\bea}{\begin{eqnarray}}
\newcommand{\eea}{\end{eqnarray}}
\newcommand{\nn}{\nonumber}
\def\be{\begin{equation}}
\def\ee{\end{equation}}
\def\ba{\begin{eqnarray}}
\def\ea{\end{eqnarray}}

\begin{document}
\baselineskip=15.5pt \pagestyle{plain} \setcounter{page}{1}

\begin{titlepage}
\begin{centering}

{\Large {\bf Backreacting $p$-wave Superconductors}}

\vspace{.3in}

Ra\'ul E. Arias $^{\dag}$\footnote{rarias@fisica.unlp.edu.ar}  and Ignacio Salazar Landea $^{\ddag\,\dag}$\footnote{peznacho@gmail.com} \\
\vspace{.2 in}
$^{\dag}${\it IFLP-CONICET and Departamento de F\'{\i}sica\\ Facultad de Ciencias Exactas, Universidad Nacional de La Plata\\
CC 67, 1900,  La Plata, Argentina}\\
~

$^{\ddag}${\it Abdus Salam International Centre for Theoretical Physics,\\
ICTP-IAEA Sandwich Training Educational Programme\\
Strada Costiera 11, 34151, Trieste, Italy}
\vspace{.4in}

{\bf Abstract} \\

\end{centering}

We study the gravitational backreaction of the non-abelian gauge field on the gravity dual to a 2+1 $p$-wave superconductor. %proposed at hep-th/08052960. %and the behavior of it's entanglement entropy.
 %In order to compare with the p+ip-wave superconductor we compute it's backreaction using an ansatz similar to those used for the p-wave gravity dual.
%In order to compare with our results for the $p$-wave superconductor we also compute the backreaction of the $p+ip$-wave superconductor.
We observe that as in the $p+ip$
system  a second order phase transition exists between a superconducting and a normal state. Moreover, %when compare the free energy
we conclude that,
%for all values of the temperature
below the phase transition temperature $T_c$ the lowest free energy %are those
is achieved by the $p$-wave solution. In order to probe the solution, we compute the holographic entanglement entropy.
For both $p$ and $p+ip$ systems the entanglement entropy satisfies an area law. For any given entangling surface, the $p$-wave superconductor has lower entanglement entropy.%values as function of the temperature for the p-wave superconductor than for the p+ip.

\end{titlepage}

\section{Introduction}

 The AdS/CFT correspondence \cite{Maldacena, Witten, Klebanov}  in its original form relates a conformal
  field theory in $d$ dimensions with type II string theory on AdS$_{d+1}$. The power of the correspondence arises from the fact
  that it is a weak/strong coupling duality {\it i.e.} it relates the strong coupling regime of the field theory with the weak coupling regime of the string theory and viceversa.
   After the works \cite{Hartnoll, Horowitz}, the gauge/gravity conjecture begun to be an useful tool to study condensed matter physics. In particular, it
    has been applied to study strongly correlated condensed matter systems through the analysis of a semiclassical dual gravity theory (see
    \cite{Iqbal:2011ae, Hartnoll:2011fn} for a review).

     In the present work we analyze
    the backreaction of the gravity dual to a $p$-wave superconductor\footnote{The tiny difference between a superconductor and a superfluid arises in the
    fact that although both effects are produced by a spontaneously symmetry breaking, in the first case there is a local symmetry which is spontaneously
    broken while in the second case is a global symmetry. We are going to use the terms superfluid and superconductor interchangably here. For the considered phenomena this distinction does not make any difference.} in 3+1 dimensions \cite{Gubserp} (see \cite{Ammon, Zhang} for a
     similar treatment in 4+1 dimensions). Along the way we rederive the backreaction of the colorful $p+ip$ superconductors previously studied in \cite{Gubserpip}.
     We use the prescription given in \cite{Ryu, Nishioka,Takayanagi} to compute the entanglement entropy from the holographic
      point of view for both gravity duals. Similar computations of entanglement entropy in backgrounds duals to condensed matter systems can be found in \cite{Albash, Cai1, Cai2, Cai3}.

$p$-wave superconductivity is a phase of matter produced when electrons with relative angular momentum $j=1$ form Copper pairs and condense.
 In other words, the operator that condense is a vector, charged under a U(1) symmetry. This kind of superconductivity is supposed to originate from "strongly
 correlated" electrons and therefore the BCS theory is not the correct approach to study its microscopic dynamics.
 This phenomena is a challenge for theoretical physics, and due to the fundamental property of the gauge/gravity duality mentioned above one could envisage the study of
 such systems through their weak gravity dual. We are going to introduce the minimal ingredients that one needs on the gravity
 side in order to reproduce the dynamics of the superconductor, this kind of approach aims to reproduce the properties of a condensed matter system without trying to explain their microscopical origin.

 String
 theory embeddings of $p$-wave superconductors were studied in \cite{Erdmenger2,Basu, Erdmenger3, Peeters, Ammon2, Kaminski}.
The necessary minimal ingredients on the bulk to have finite temperature, chemical potential and spontaneous symmetry breaking (SSB) are:
a black hole geometry and a non-Abelian gauge field \cite{Hartnoll2, Hartnoll3}. The solutions we will consider are asymptotically AdS backgrounds with a $SU(2)$ non-Abelian gauge field. The SSB is realized on the bulk side as a non-trivial asymptotics (hair)
for the gauge field. The chemical potential and the SSB arise by turning on two independent directions inside the non-Abelian gauge group.
 The symmetry breaking occurs on the gravity side through the formation of a condensate outsides the horizon.

The entanglement entropy (EE) between a subsystem ${\cal A}$ and it's
complement ${\cal B}$ is the von Neumann entropy
\be
{\cal S_{\cal A}}=-Tr_{{\cal A}}(\rho_{\cal A}\ln\rho_{\cal A}).
\ee
Here $\rho_{\cal A}=Tr_{\cal B}(\rho)$ is the density matrix obtained by tracing the density matrix of the whole system $\rho$ over the ${\cal B}$ subsystem degrees of freedom. Roughly speaking $S_{\cal A}$ measures how much information is hidden
inside ${\cal B}$ when we subdivide the system.
From the point of view of the dual gravity theory the EE was conjectured \cite{Ryu} to be proportional to the bulk minimal
area surface, $\gamma_{\cal A}$, whose boundary at infinity coincides with the boundary of ${\cal A}$ (see \cite{Takayanagi} for a review)
\be
{\cal S_{\cal A}}= \frac{2\pi Area(\gamma_{\cal A})}{\kappa^2}.
\ee
Here $\kappa$ is the bulk gravitational constant. Note that the standard thermal entropy is obtained as a particular case of the EE, when the region ${\cal A}$
is the whole system.
In \cite{Casini} the authors provide a demonstration of this holographic technique to compute the entanglement entropy for
spherical surfaces and zero temperature CFTs.
On this work we compute this quantity for a strip geometry in the backgrounds dual to a $p$-wave and to a colorful $p+ip$ superconductor.

This paper is organized as follows: in section \ref{holsup} we compute the backreaction of a 3+1 gravity dual to a $p$-wave superconductor in 2+1 dimensions and analyze its thermodynamic
properties. In order to compare with the colorful superconductor, we review in subsection \ref{p+ip} the backreaction of the gravity dual of a
$p+ip$ superconductor. %written the background ansatz in a similar form to those used in subsection \ref{pwave}.
On section \ref{Entanglement}
we compute the holographic entanglement entropy for a strip geometry in both systems, as a function of the temperature and the length of the strip.
The conclusions are summarized in section \ref{summary}.

\section{$p$ and $p+ip$ holographic superconductors}\label{holsup}

As mentioned in the introduction, the gravity dual to a $p$-wave superconductor is modeled by  an Einstein-Yang-Mills (EYM) theory. In \cite{Gubserp, Gango}, the 3+1 dimensional gravity theory dual to a $p$-wave superconductor  has been computed in the probe limit. Moreover the authors showed
that the $p+ip$ superconductor geometry studied in \cite{Gubserpip} was unstable under small fluctuations, and that the stable configuration was that of the $p$-wave solution.
In this section we compute the backreaction of the non-Abelian gauge field on the geometry dual to a $p$-wave superconductor in 3+1 dimensions and compare the results with those for the $p+ip$ case.

 We will work in the simplest set up and consider $SU(2)$ as the gauge group. In the $p+ip$ case, the ansatz for the gauge
field is such that it breaks the U(1) subgroup of the internal
gauge SU(2) and the spatial rotational SO(3) group symmetries into
a diagonal subgroup of them. Instead, the $p$-wave superconductor,
breaks both U(1) symmetries completely. The gravity solution that
describes the strong coupling dynamics of both kinds of
superconductors is as follows: a charged superconducting layer
develops outside the horizon due to the interplay between the
electric repulsion (with the charged black hole) and the
gravitational potential of the asymptotically AdS geometry. At
high enough temperatures there is no hair outside the black hole
and the solution is just an  AdS-Reissner-Nordström (AdSRN) black
hole. Below a critical temperature $T_c$ a non-trivial gauge field
with non-vanishing chemical potential on the boundary of the
geometry and a sourceless non-vanishing condensate in the bulk
appears, originating a breaking of the SU(2) gauge symmetry.

%The p+ip-wave superconductor studied on \cite{Gubserpip} has a second order phase transition and as was showed in \cite{Gubserp} it is an unstable
%solution, under perturbations that break rotational invariance, in the limit of large Yang-Mills coupling constant (i.e. the probe limit)
%and near the critical temperature. Here the ip component
%decreases and appears likely to lead the system into a p-wave state. Then we expect to find a second order phase transition for the p-wave %superconductor
 %on 3+1 dimensions.

%The dual conformal field
%theory has a global SU(2) current algebra. In the small coupling limit the picture is that the condensate is formed by a pair of fermions which have angular momentum $l=1$.

\subsection{$p$-wave superconductor in 2+1 dimensions}\label{pwave}
\subsubsection{Solution}
We start from $3+1$ $SU(2)$ Yang Mills Theory in AdS gravity (see \cite{Winst} for a review about solutions for this theory), the Lagrangian density is
\be
\label{action}
\kappa_{(4)}^2{\cal L} =R -2\Lambda-\frac14Tr(F_{\mu\nu}F^{\mu\nu})%= R +\frac{6}{L^2}- \frac14Tr(F_{\mu\nu}F^{\mu\nu})
\ee
where $\Lambda=-\frac{3}{{\hat R}^2}$, $\kappa_{(4)}$ is the gravitational constant in four dimensions and the field strength of the SU(2) gauge field is written as
\begin{equation}
\label{energymomentumtensor}
F^a_{\mu \nu}=\partial_\mu A^a_\nu-\partial_\nu A^a_\mu+g_{_{YM}}\epsilon^{abc}A^b_\mu A^c_\nu
\end{equation}
with $g_{_{YM}}=\frac{\hat g_{_{YM}}}{\kappa_{(4)}}$ the parameter that measures the backreaction and $\hat g_{_{YM}}$ the usual Yang-Mills coupling. We use latin letters for SU(2) indexes and
greek letters for the space-time coordinates. By scaling the gauge field as $\tilde A= \frac{A}{g_{_{YM}}}$
we see that the large $g_{_{YM}}$ limit corresponds to the probe (non-backreacting) limit of the gauge field.
Roughly one can think that $\frac{1}{\hat g_{_{YM}}^2}$ counts the degrees of freedom of the dual field theory that are charged under the $SU(2)$ gauge
group. Moreover, $\frac{1}{\kappa_{(4)}^2}$ counts the total number of degrees of freedom. Considering backreaction of the gauge filed amounts to say that the number of charged states is of the same order as the number of degrees of freedom of the system.
%because in general the correlation functions of $SU(2)$ currents are proportional to $\frac{1}{\hat g_{_{YM}}^2}$.
%Using the following definition for the energy-momentum tensor:
%\be
%T_{\mu\nu}=-\frac{1}{\sqrt{-g}}\frac{\delta {\cal L}}{\delta g^{\mu\nu}}
%\ee

The equations of motion following from the action are
\bea\label{EYM}
G_{\mu\nu}= R_{\mu\nu}-\frac12g_{\mu\nu} R&=&\frac{3}{R^2} g_{\mu\nu}+\frac12Tr[F_{\mu\gamma}F_\nu^\gamma]-\frac{g_{\mu\nu}}{8}Tr[F_{\gamma\rho}F^{\gamma\rho}]\\
\nonumber
\\
D_\mu F^{\mu\nu}&=&0\label{Max}
\eea
we propose the following ansatz \cite{Ammon, Manvelyan}
\begin{equation}
\label{metricansatz}
ds^2 = -M(r)\sigma(r)^2dt^2 + \frac{1}{M(r)}dr^2 +r^2 h(r)^{2}dx^2 + r^2h(r)^{-2} dy^2 \,,
\end{equation}
for the background geometry, and
\begin{equation}
\label{gaugefieldansatz}
A=\phi(r)\tau^3 dt+\omega(r)\tau^1d x\,.
\end{equation}
for the gauge field. Here we use the matrix-valued notation $A=A_\mu^a\tau^adx^\mu$ with $\tau^a=\frac{\sigma^a}{2i}$ and $\sigma^a$ the usual Pauli matrices, the $SU(2)$ generators satisfy $[\tau^a,\tau^b]=\epsilon^{abc}\tau^c$. A solution developing $\omega\neq0$ in the gauge field ansatz \eqref{gaugefieldansatz} breaks the $U(1)$ gauge symmetry associated with rotations around $\tau^3$ (usually called $U(1)_3$)
and a $h\neq0$ in the metric breaks $U(1)_{xy}$ symmetry associated to rotations on the $xy$ plane. At high enough temperatures we expect no hair outside the black hole and the solution with no condensate is AdSRN with
\bea
\omega(r)&=&0,\,\,\,\nn\\  h(r)&=&1 \nn\\\sigma(r)&=&1,\,\,\, \nn\\ \phi(r)&=&\mu\left(1-\frac{r_h}{r}\right),\nn\\ M(r)&=&r^2+\frac{\mu^2r_h^2}{r^2}-\left(\frac{\mu^2}{8}+r_h^2\right)\frac{r_h}{r}.
\eea

Replacing the ansatz into the EYM equations of motion results into
five equations, three of them are second order differential
equations, and the remaining two are first order constraints \bea
M'&=&\frac{3r}{{\hat
R}^2}-\frac{1}{8\sigma^2}\left(\frac{g_{_{YM}}^2\phi^2\omega^2}{rh^2M}+r\phi'^2\right)-M
\left(\frac{1}{r}+\frac{rh'^2}{h^2}+\frac{\omega'^2}{8rh^2}\right)\nn\\
\sigma'&=&\frac{\sigma}{h^2}\left(r
h'^2+\frac{\omega'^2}{8r}\right)+\frac{g_{_{YM}}^2\phi^2\omega^2}{8r
M^2h^2\sigma};\nn\\
h''&=&\frac{1}{8r^2h}\left(-\omega'^2+\frac{g_{_{YM}}^2\phi^2\omega^2}{M^2\sigma^2}\right)-h'\left(\frac{2}{r}-\frac{h'}{h}+\frac{M'}{M}+\frac{\sigma'}{\sigma}
\right);\nn\\
\omega''&=&-\frac{g_{_{YM}}^2\phi^2\omega}{M^2\sigma^2}+\omega'\left(\frac{2h'}{h}-\frac{M'}{M}-\frac{\sigma'}{\sigma}\right);\nn\\
\phi''&=&\frac{g_{_{YM}}^2\phi\,\omega^2}{r^2h^2M}-\phi'\left(\frac{2}{r}-\frac{\sigma'}{\sigma}\right).
\label{eomsp}
\eea
This system of equations enjoys four scaling symmetries that become useful when numerically solving it, they are
\begin{enumerate}
    \item $\sigma\rightarrow\lambda\sigma,~~~~\phi\rightarrow\lambda\phi$
    \item $\omega\rightarrow\lambda\omega,~~~~h\rightarrow\lambda h$
    \item $M\rightarrow\lambda^{-2} M,~~~~\sigma\rightarrow\lambda \sigma,~~~~g_{_{YM}}\rightarrow\lambda^{-1}g_{_{YM}},~~~~{\hat R}\rightarrow\lambda {\hat R}$
    \item $M\rightarrow\lambda^{2} M,~~~~ r\rightarrow \lambda r,~~~~\phi\rightarrow\lambda\phi,~~~~\omega\rightarrow\lambda\omega $
\end{enumerate}
Using these scaling symmetries we can set $R=r_h=1$ and fix the boundary value of the metric functions $\sigma(\infty)=h(\infty)=1$. The geometry and the gauge field must be regular at the horizon which implies the following expansion in the IR (small $r$)
\bea\label{IRp}
M&=&M_1(r-r_h)+M_2(r-r_h)^2+\ldots\nn\\
h&=&h_0+h_2(r-r_h)^2+\ldots\nn\\
\sigma&=&\sigma_0+\sigma_1(r-r_h)+\sigma_2(r-r_h)^2+\ldots\nn\\
\omega&=&\omega_0+\omega_2(r-r_h)^2+\omega_3(r-r_h)^3+\ldots\nn\\
\phi&=&\phi_1(r-r_h)+\phi_2(r-r_h)^2+\ldots
\eea
On other hand in the UV (large $r$) the desired behavior is:
\bea
M&=& r^2+\frac{M_1^b}{r}+\frac{(\omega_1^b)^2+\rho^2}{8r^2}+\ldots\nn\\
h&=&1+\frac{h_3^b}{r^3}-\frac{(\omega_1^b)^2}{32r^4}+\ldots\nn\\
\sigma&=&1-\frac{(\omega_1^b)^2}{32r^4}+\ldots\nn\\
\omega&=&\omega_0^b+\frac{\omega_1^b}{r}-\frac{g_{_{YM}}^2\mu^2\omega_1^b}{6r^3}+\ldots\nn\\
\phi&=&\mu+\frac{\rho}{r}+\frac{g_{_{YM}}^2\mu^2\omega_1^b}{12r^4}+\ldots\label{bcp}
\eea
To achieve SSB, we look for solutions where the non-normalizable component vanishes $\omega_0^b=0$. Standard AdS/CFT dictionary instruct us to interpret
the boundary and sub-leading values of $\phi$ as the chemical potential $\mu$ and the charge density $\rho$ of the dual field theory \cite{KW}. Moreover,
the sub-leading coefficient $M_1^b$ in the boundary expansion of $g_{tt}$ coincides with the regularized Euclidean on-shell action \cite{Hartnoll}. The normalizable coefficient in $\omega$ is dual to the vacuum expectation value of the current $\langle J_x^1\rangle\propto \omega_1^b$ and serves as an order parameter for the system.

\begin{figure}[ht]
\begin{minipage}{8cm}
\begin{center}
\includegraphics[width=8cm]{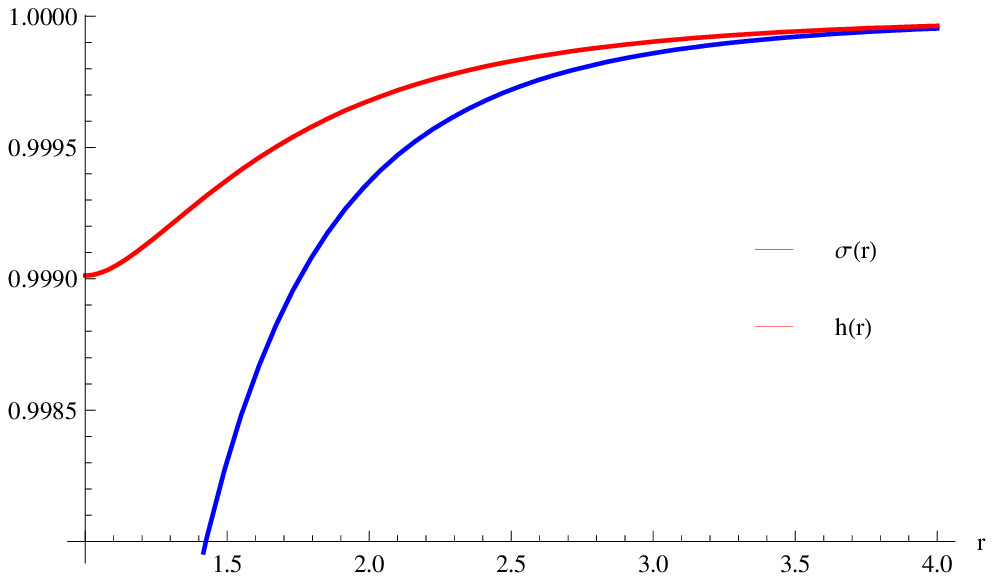}
\caption{The dimensionless metric functions $\sigma(r)$ and $h(r)$ for $g_{_{YM}}=2$ and $T=0.2312\mu$.}
\label{hysigmap}
\end{center}
\end{minipage}
\  \
\hfill \begin{minipage}{8cm}
\begin{center}
\includegraphics[width=8cm]{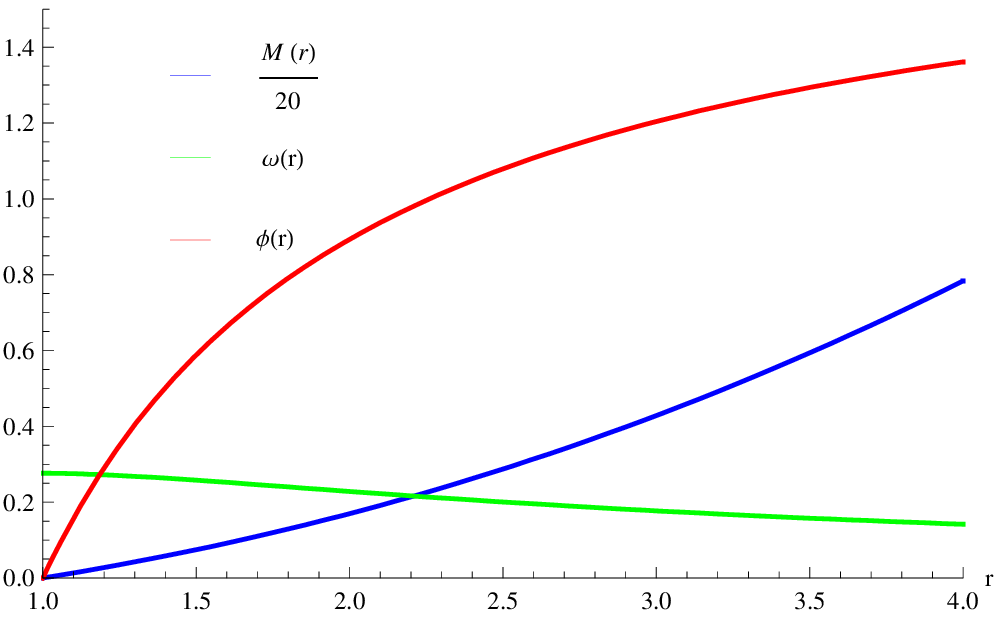}
\caption{The dimensionless metric function $M(r)$ and the gauge field functions $\omega(r)$ and $\phi(r)$ for $g_{_{YM}}=2, T=0.2312\mu$.}
\label{funcionesp}
\end{center}
\end{minipage}
\end{figure}

Solutions of the system \eqref{eomsp} depend on the four IR coefficients $\phi_1,\omega_0,h_0,\sigma_0$ and the backreaction parameter $g_{_{YM}}$. All other coefficients in \eqref{IRp} can be written in terms of those.
We proceed to integrate the equations of motion numerically out from the horizon using a shooting method in order to get the desired asymptotic behavior. We explore the range $g_{_{YM}}\in [0.85,24]$ and observe that the behavior of the functions does not change qualitatively as $g_{_{YM}}$ is varied.
In figure \ref{hysigmap} and \ref{funcionesp} we give the plot of the solutions of \eqref{eomsp} with boundary conditions \eqref{bcp}. %for the value $g_{_{YM}}=1$
 %and temperature $T=0.0749\mu$.
We use $\mu$ to adimensionalize whenever needed. This means that we are working in the grand canonical ensemble.

\subsubsection{Thermodynamics}

In this section we compute the thermodynamic quantities associated with the solutions. As we shall see
from the study of the potential function in the grand canonical ensemble\footnote{To go from the
grand canonical ensemble (fixed $\mu$) with free energy $\Omega$, to the canonical ensemble (fixed $\rho$) with free energy $F$, we should add a boundary term to the Euclidean action. This changes
the variational problem and implies the known Gibbs relation $F=\Omega+\mu\rho$.}
 we have a second order phase transition between a superconducting and normal symmetric phases.

\begin{figure}[h]
%\begin{minipage}{7.7cm}
\begin{center}
\includegraphics[width=7.7cm]{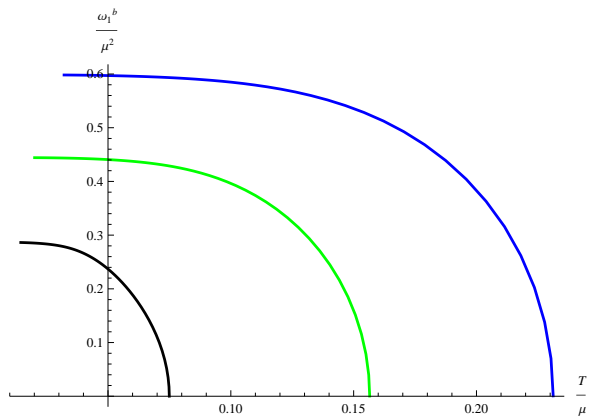}
\caption{The plot shows the normalizable coefficient of the $\omega$ function which is proportional to the condensate $\langle J_x^1\rangle$. The black, green and blue lines refers to solutions with $g_{_{YM}}=1, 1.5, 2$ and $T_c=0.0749, 0.1565, 0.2312$ respectively. Note that the condensate vanishes for $T>T_c$.}
\label{condp}
\end{center}
%\end{minipage}
\end{figure}

\begin{figure}[h]
\begin{minipage}{7.7cm}
\begin{center}
\includegraphics[width=7.7cm]{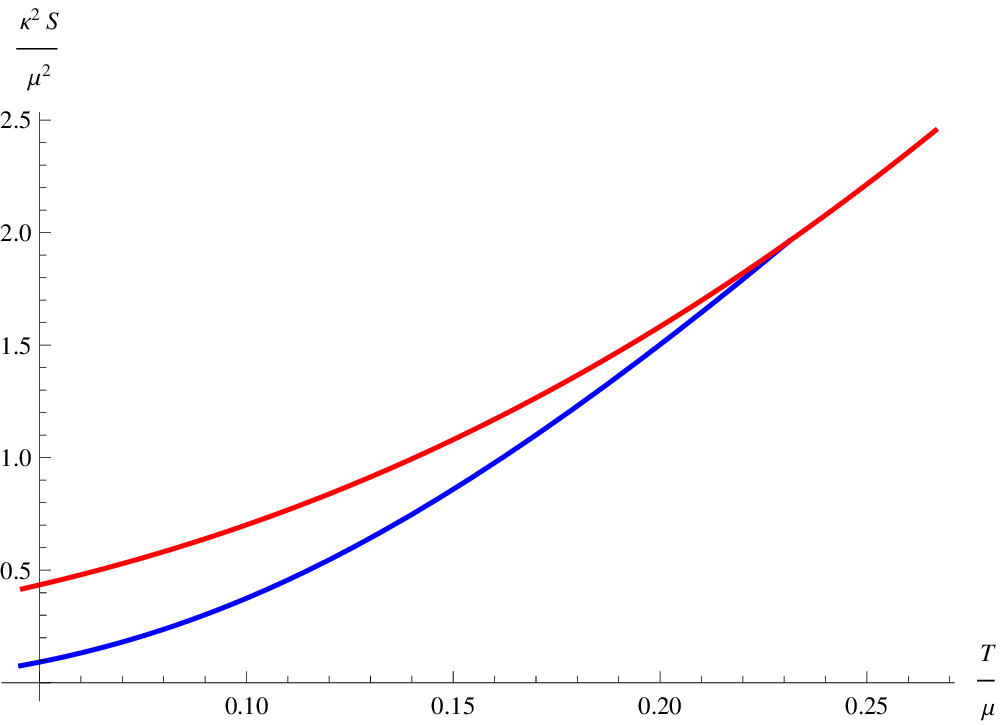}
\caption{The entropy as a function of the temperature. The blue line is for the superconducting phase with $g_{_{YM}}=2$ and the red line for the normal phase (AdSRN geometry). There is a second order phase transition at $T=T_c=0.2312$. }
\label{EvsTp}
\end{center}
\end{minipage}
\ \
\hfill\begin{minipage}{7.7cm}
\begin{center}
\includegraphics[width=7.7cm]{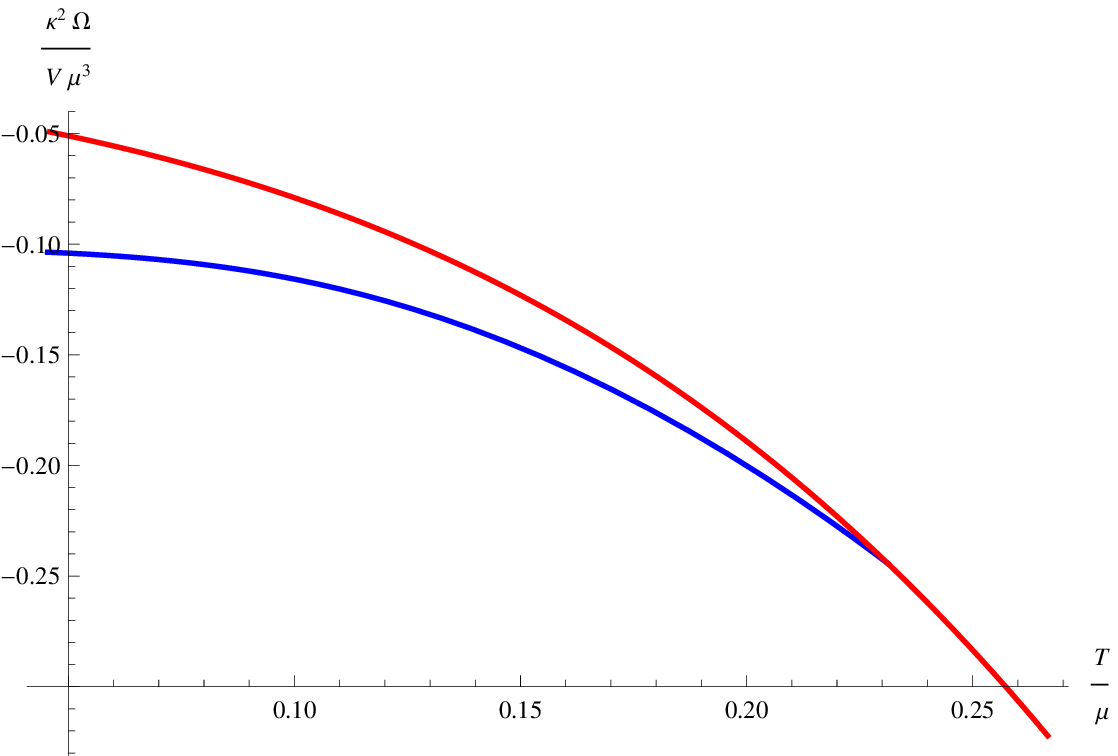}
\caption{The potential function $\Omega$ computed from \eqref{omega} as a function of $T$ for $g_{_{YM}}=2$.
The red line is the potential for the RN solution and the blue line is for the superconductor case. }
\label{freep}
\end{center}
\end{minipage}
\end{figure}

The temperature of the dual theory is given by the Hawking temperature of the black hole
\be
T=\frac{M_1\sigma_0}{2\pi}=\frac{1}{16\pi\sigma_0}\left(24\sigma_0^2-\phi_1^2\right)r_h
\ee
where the second equality comes from the consistency of the series expansion \eqref{IRp} that relates the
coefficient $M_1$ with $\sigma_0$ and $\phi_1$. The area of the horizon, $A_h$, gives the entropy
\be
S=\frac{2\pi}{\kappa^2_{(4)}} A_h%=\frac{2\pi}{\kappa^2_{(4)}} r_h^2
=\frac{2\pi^2VT^2}{\kappa^2_{(4)}}\frac{12^2}{\left(24\sigma_0^2-\phi_1^2\right)^2}\label{Sp}
\ee
where $V=\int dx\,dy$. In figure \ref{condp} we plot the order parameter $\omega_1^b$(i.e. the VEV of the current $\langle J_x^1\rangle$) as a function of the temperature.
Note that at $T=T_c$ the condensate vanishes showing the disappearance of the superconducting state for $T>T_c$. From our
numerical results we find  $\langle J_1^x\rangle\propto(1-\frac{T}{T_c})^{1/2}$ near $T_c$ and therefore the critical exponent takes the value $1/2$.
In reference \cite{Ammon} the authors scan the range $g_{_{YM}}\in [1.82,31.5]$ and find that the phase transition becomes first order for $g_{_{YM}}<2.74$. We didn't find such first order phase transition in the $3+1$ case.
Figure \ref{EvsTp} shows the behavior of the Bekenstein-Hawking entropy \eqref{Sp} as function of the temperature for our solution and the AdSRN black hole.

The gauge/gravity correspondence identifies the Euclidean on-shell gravity action $S_E$ times the temperature $T$ as the grand canonical potential function $\Omega$ of the system. To compute it we continue to Euclidean signature, time being compactified with period $\frac{1}{T}$ to avoid singularities. The on-shell action has a factor $\frac{1}{T}$ due to time integration, writing $ S_{on-shell}=\frac{\tilde S_{bulk}}{T}$ one has
\be\label{onshellp}
\tilde S_{bulk}=-\int dx\,dy\,dr\sqrt{-g}\,{\cal L}
\ee
where the lagrangian density is given by \eqref{action}. The $yy$ component of the stress tensor is proportional to the metric and then the Einstein equations \eqref{EYM} implie that
\be
G_{yy}=\frac{r^2}{2h^2}\left(\kappa^2_{(4)}{\cal L}-R\right)
\ee
Then we have
\be
G_\mu^\mu=-R=G_r^r+G_t^t+G_x^x+\frac{1}{2}\left(\kappa^2_{(4)}{\cal L}-R\right)
\ee
and from this we obtain
\be
{\cal L}=\frac{2}{r^2\sigma\kappa^2_{(4)}}\left[\frac{r^3M\sigma}{h}\left(\frac{h}{r}\right)'\right]'
\ee
where $'$ denotes derivative with respect to the holographic coordinate $r$. Then, the bulk contribution to the
on shell action \eqref{onshellp} can be written as
\be
\tilde S_{bulk}=-\int dx\,dy\,dr\sqrt{-g}{\cal L}=-\frac{2V}{\kappa^2_{(4)}} \left[\frac{r^3 M\sigma}{h}\left(\frac{h}{r}\right)'\right]_{r=r_\infty}
\ee
where $r_{\infty}$ is the boundary of the space.  As usual, in order to have a well defined variational problem when
imposing Dirichlet boundary conditions on the metric we need to add to the action a Gibbons-Hawking term
\be
\tilde S_{GH}=-\frac{1}{\kappa^2_{(4)}}\int dx\,dy\sqrt{-g_{\infty}}\,\nabla_\mu n^\mu=-\frac{V}{\kappa^2_{(4)}}r^2\sigma\left[\frac{M'}{2}+M\left(\frac{\sigma'}
{\sigma}+\frac{2}{r}\right)\right]_{r=r_\infty},\label{GH}
\ee
where $n^\mu dx_\mu=\sqrt{M}dr$ is the outward pointing unit normal vector to the boundary and $g_\infty$ is the determinant of the induced
metric on the boundary. Precisely at $r=r_{\infty}$ \eqref{GH} diverges and therefore must be regularized adding the intrinsic boundary counter-term
\be
\tilde S_{ct}=\frac{1}{\kappa^2_{(4)}}\int dx\,dy\sqrt{-g_{\infty}}=\frac{V}{\kappa^2_{(4)}}\left[r^2\sqrt{M}\sigma\right]_{r=r_\infty}
\ee
 Finally the dual thermodynamic potential $\Omega$ results
\bea\label{omega}
\Omega&=&\lim_{r_\infty\rightarrow\infty}\tilde S_{on-shell}\nn\\
&=&\lim_{r_\infty\rightarrow\infty}(\tilde S_{bulk}+\tilde S_{GH}+\tilde S_{ct})
\eea

Upon regularizing the action the potential $\Omega$ results to coincide with the sub-leading value of the $g_{tt}$ component of the
background metric i.e. $\Omega=M_1^b$ \cite{Hartnoll}. We have verified our numerical solution computing $\Omega$ in both ways finding an excellent agreement.
In figure \ref{freep} we plot the potential \eqref{omega} as function of the temperature.
%for $g_{_{YM}}=2$ (we explore the range of values between $0.85$ and $24$ and the behavior of the functions is qualitatively the same for all of them).
As we mentioned above a second order phase transition develops at $T=T_c$: the grand potential and the entropy are continuous but $S$ is not differentiable. Below $T_c$ the system is in the superconducting phase,
as we increase the temperature above $T_c$ the AdSRN geometry dominates the free energy, this models a transition from
a superconducting to a normal phase.

\subsection{$p+ip$ wave superconductors}\label{p+ip}

Here, we review the results of \cite{Gubserpip} and compare them with the results of the previous section.
We will find that at $T=T_c$ the system has a second order phase transition and for all ranges of temperatures the grand potential of the $p$-wave solution found in previous section is lower than that of the $p+ip$, implying  that the stable phase of the system is the $p$-wave phase in accordance with the stability analysis \cite{Gubserp}.

\subsubsection{Solution}

The background and gauge field ansatz for model a $p+ip$-wave solution are
\bea
ds^2&=&-M(r)dt^2+r^2h(r)^2(dx^2+dy^2)+\frac{dr^2}{M(r)}\\
A&=&\phi(r)\tau^3dt+\omega(r)(\tau^1dx+\tau^2dy).
\eea
One important difference with the $p$-wave superconductor of the previous section arises in the choice of the gauge field ansatz
that now breaks the $U(1)_3\times U(1)_{xy}$ into a diagonal combination. The $p$-wave case fully breaks the $U(1)_3\times U(1)_{xy}$. This allows us to use a metric ansatz that is totationally symmetric in the $xy$-plane.

\begin{figure}[ht]
\begin{minipage}{7.8cm}
\begin{center}
\includegraphics[width=7.8cm]{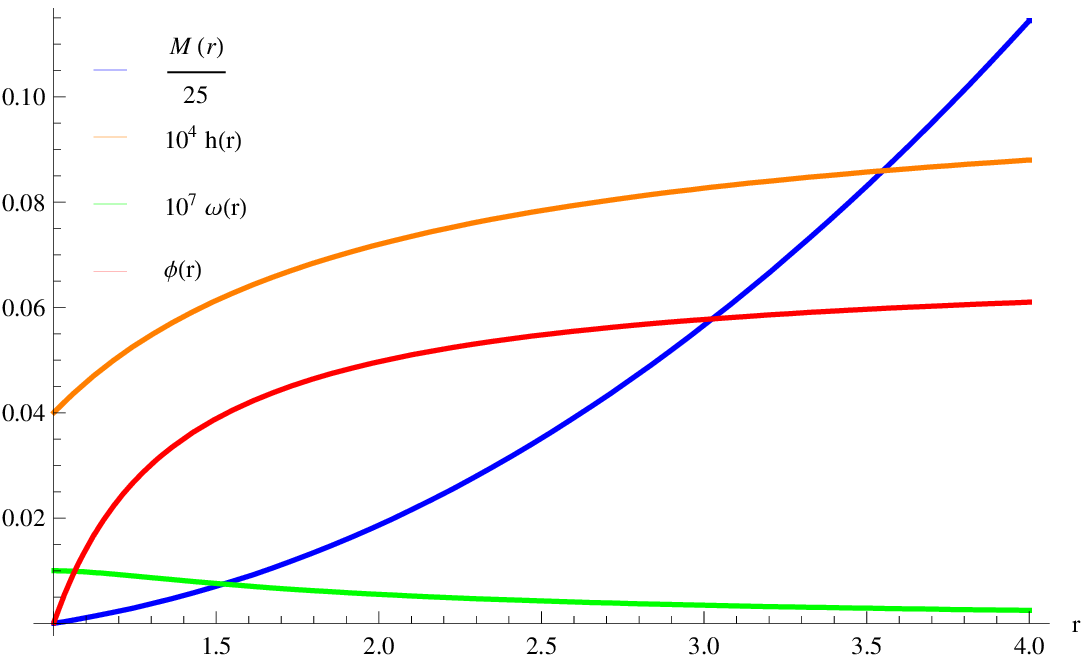}
\caption{Behavior of the dimensionless functions $M(r),h(r),\omega(r)$ and $\phi(r)$, plotted for $g_{_{YM}}=2, T=0.2312\mu$.}
\label{funcionesp+ip}
\end{center}
\end{minipage}
\ \ \ \
\hfill\begin{minipage}{8cm}
\begin{center}
\includegraphics[width=8cm]{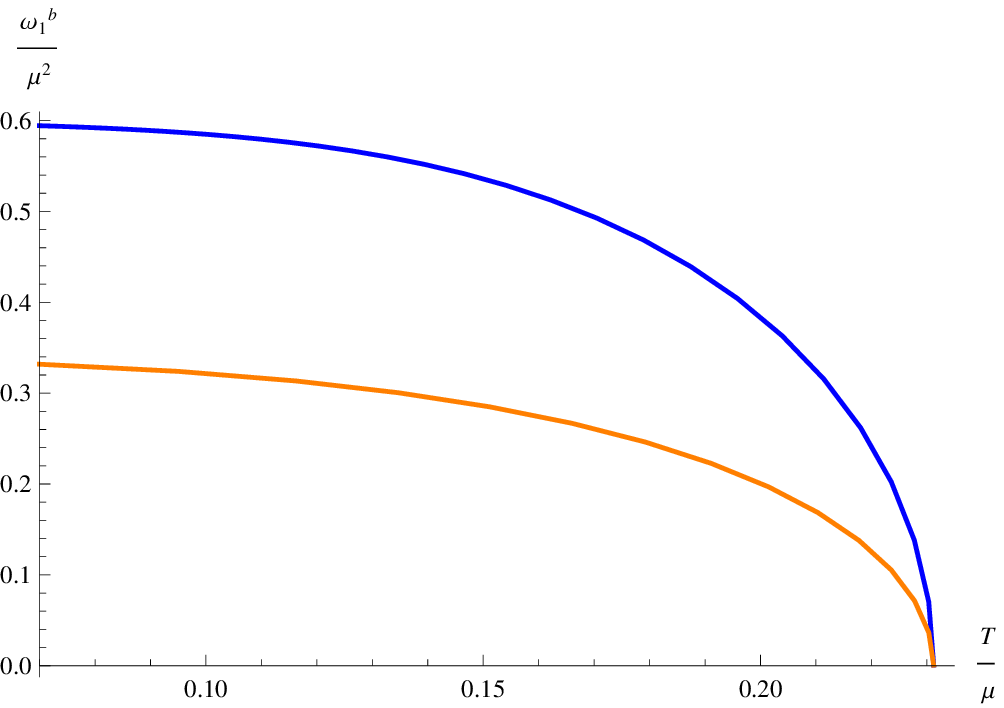}
\caption{The dual theory VEV $\langle J_x^1\rangle\propto \omega_1^b$ as a function of temperature for the case of the p-wave (blue line) and colorful
(orange line) superconductors for $g_{_{YM}}=2$. Its vanishing for $T>T_c=0.2312$,  suggesting a phase transition between
a superconducting and a normal state.}
\label{condpypip}
\end{center}
\end{minipage}
\end{figure}

The equations of motion obtained for this ansatz are four second order differential equation plus a first order constraint arising from the $rr$ component of the Einstein equations
\bea
h''&=&-\frac{h}{2}\left[\frac{1}{r^2}-\frac{3}{{\hat R}^2M}+\frac{M'}{r M}+\frac{\phi'^2}{8M}+\frac{\omega'^2}{4r^2h^2}\right]-\frac{h'}{2}\left[\frac{6}{r}+\frac{h'}{h}
+\frac{M'}{M}\right]-\frac{g_{_{YM}}^2\omega^2}{8r^2hM}\left[\frac{\phi^2}{M}+\frac{\omega^2}{2r^2h^2}\right]\nn\\
M''&=&\frac{3}{{\hat R}^2}+\frac{M}{r}\left[-\frac{M'}{M}+\frac1r+\frac{\omega'^2}{4rh^2}\right]-\frac{h'}{h}\left[M'-\frac{h'}{h}-\frac2r\right]+\frac38\phi'^2+
\frac{g_{_{YM}}^2\omega^2}{4r^2h^2}\left[\frac{\phi^2}{M}+\frac{3\omega^2}{2r^2h^2}\right]\nn\\
\omega''&=&\frac{g_{_{YM}}^2\omega}{M}\left[\frac{\omega^2}{r^2h^2}-\frac{\phi^2}{M}\right]-\frac{M'\omega'}{M}\nn\\
\phi''&=&\frac{2g_{_{YM}}^2\phi\omega^2}{r^2h^2M}-2\phi'\left[\frac1r+\frac{h'}{h}\right]\nn\\
0&=&-\frac{3}{{\hat R}^2}+\frac{M}{r^2}\left[1-\frac{\omega'^2}{4h^2}+\frac{M'}{M}r\right]+\frac{h'}{h}\left[M\left(\frac2r+\frac{h'}{h}\right)
+M'\right]+\frac18\phi'^2.\nn
\eea\label{pip}
The equations have three scaling symmetries that will help us to numerically solve the system. They are
\begin{enumerate}
\item $\omega\rightarrow\lambda\omega,~~~~h\rightarrow\lambda h$
\item $M\rightarrow\lambda^{-2}M,~~~~\phi\rightarrow\frac{\phi}{\lambda},~~~~{\hat R}\rightarrow\lambda {\hat R},~~~~g_{_{YM}}\rightarrow\frac{g_{_{YM}}}{\lambda}$
\item $M\rightarrow\lambda^2M,~~~~h\rightarrow\frac{h}{\lambda},~~~~\phi\rightarrow\lambda\phi,~~~~r\rightarrow\lambda r$\label{scalpip}
\end{enumerate}
and allows us to set $R=r_h=1$ and the value of $h(r)$ at the boundary to $h(\infty)=1$. The IR behavior of these equations are those
of a charged black hole
\bea\label{IRpip}
M&=&M_1(r-r_h)+M_2(r-r_h)^2+\ldots\nn\\
h&=&h_0+h_1(r-r_h)+h_2(r-r_h)^2+\ldots\nn\\
\omega&=&\omega_0+\omega_1(r-r_h)+\omega_2(r-r_h)^2+\ldots\nn\\
\phi&=&\phi_1(r-r_h)+\phi_2(r-r_h)^2+\ldots
\eea
where as before we impose the Maxwell potential $\phi$ to vanish at the horizon in order to have a well defined gauge field in the Euclidean continuation.
On the UV we demand
\bea
M&=&r^2+2 h_1^b r+(h_1^b)^2+\frac{M_1^b}{r}+\frac{-8h_1^b M_1^b+\rho^2+2(\omega_1^b)/3}{8r^2}+\ldots\nn\\
h&=&1+\frac{h_1^b}{r}-\frac{(\omega_1^b)^2}{48 r^4}+\ldots\nn\\
\omega&=&\frac{\omega_1^b}{r}-\frac{h_1^b\omega_1^b}{r^2}+\ldots\nn\\
\phi&=&\mu+\frac{\rho}{r}-\frac{\rho h_1^b}{r^2}+\ldots
\eea
%\bea
%M&=&r^2+M_0^b+\frac{M_1^b}{r}\ldots\nn\\
%h&=&h_0^b+\frac{h_1^b}{r}+\frac{h_4^b}{r^4}+\ldots\nn\\
%\omega&=&\omega_0^b+\frac{\omega_1^b}{r}+\frac{\omega_2^b}%{r^2}+\frac{\omega_3^b}{r^3}+\ldots\nn\\
%\phi&=&\phi_0^b+\frac{\phi_1^b}{r}+\frac{\phi_2^b}%{r^2}+\frac{\phi_3^b}{r^3}+\ldots
%\eea
Note that for SSB we do not allow for a non-normalizable piece in
$\omega$. As before the scaling symmetry \eqref{scalpip} allows to
fix $h_0^b=1$. In figure \ref{funcionesp+ip} we plot the behavior
of the solutions and
 figure \ref{condpypip} shows the order parameter $\langle J_x^1\rangle>\propto \omega_1^b$ as function of the temperature.
For $T=T_c$ both condensates vanish and a second order phase transition onsets.
 Note that the values of the condensate for the $p+ip$ case are lower than those in the $p$-wave case.

\subsubsection{Thermodynamics}

\begin{figure}[ht]
\begin{minipage}{8cm}
\begin{center}
\includegraphics[width=7.5cm]{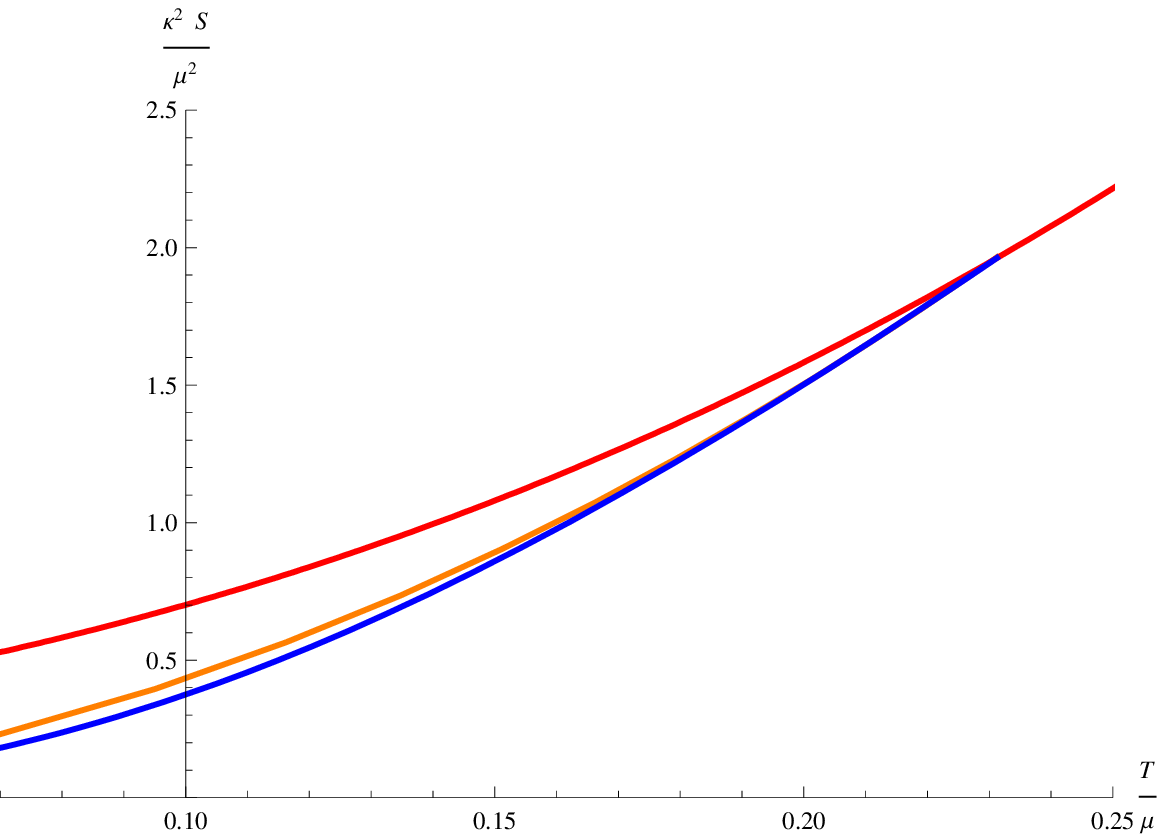}
\caption{Bekenstein-Hawking entropy for the RN (in red), $p+ip$ (orange) and $p$-wave (blue) solutions with $g_{_{YM}}=2$ and $T_c=0.2312$. There is a second order phase transition for both superconductors at $T=T_c$. }
\label{SvsT}
\end{center}
\end{minipage}
\ \
\hfill\begin{minipage}{8cm}
\begin{center}
\includegraphics[width=7.5cm]{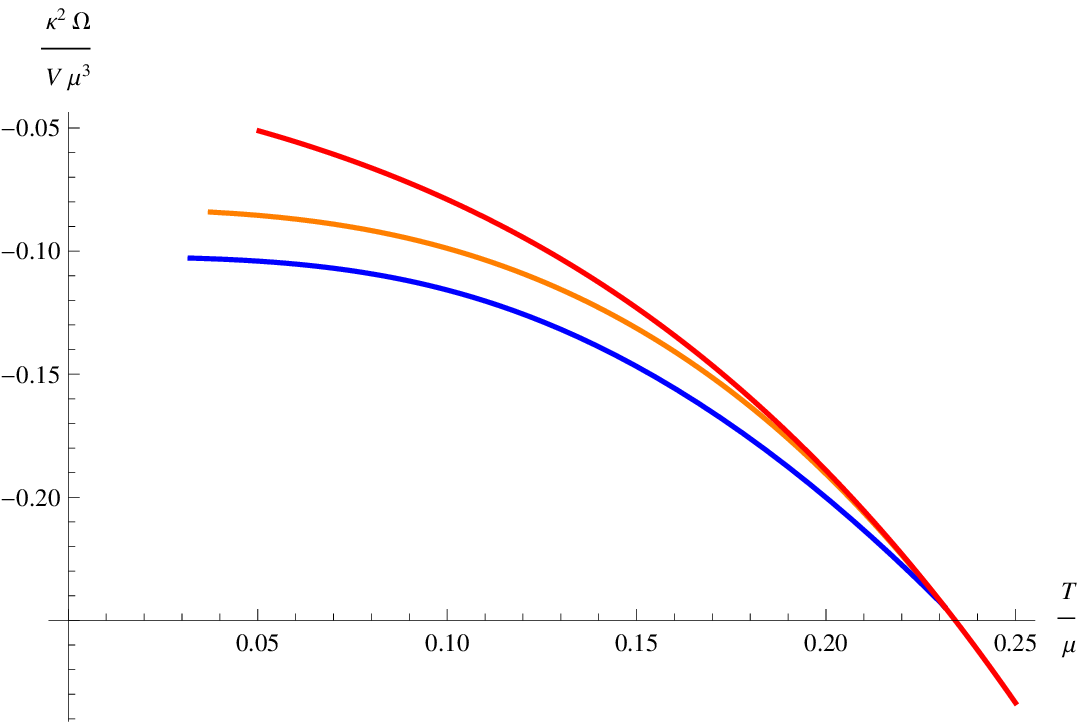}
\caption{Grand Canonical potential as a function of the temperature. RN solution (red), $p+ip$ (orange)
and $p$-wave (blue). For all range of temperatures below $T_c$ the $p$-wave solution is preferred over the colorful one ($g_{_{YM}}=2$ and $T_c=0.2312$). }
\label{FvsT}
\end{center}
\end{minipage}
\end{figure}

The temperature associated to the background solution is proportional to the derivative of the $g_{tt}$ component of the metric evaluated at the horizon.
In this case one has
\be
T=\frac{M_1}{2\pi}
\ee
The Bekenstein-Hawking formula, that relates the entropy with the area of the black hole horizon, for the present case reads
\be
S=\frac{2\pi}{\kappa^2_{(4)}} A_h=\frac{2\pi}{\kappa^2_{(4)}} r_h^2h_0^2.
\ee
Figure \ref{SvsT} shows the entropy in the $p+ip$ (orange line), $p$-wave (blue line) and RN (red line) cases.

The grand potential $\Omega$, is given by the sub-leading coefficient of the metric function $g_{tt}$, as
\be
\Omega=\frac{V M_1^b}{\kappa^2_{(4)}}
\ee
This potential is plotted in figure \ref{FvsT}, it clearly shows that for any given temperature the $p$-wave solution (blue) is preferred over the $p+ip$ (orange) state.
For $T>T_c$ the system is in the normal phase (red) and the condensate (shown in fig. \ref{condpypip}) vanishes.

\section{Holographic Entanglement Entropy}\label{Entanglement}

\begin{figure}[ht]
%\begin{minipage}{7cm}
\begin{center}
\includegraphics[width=10cm]{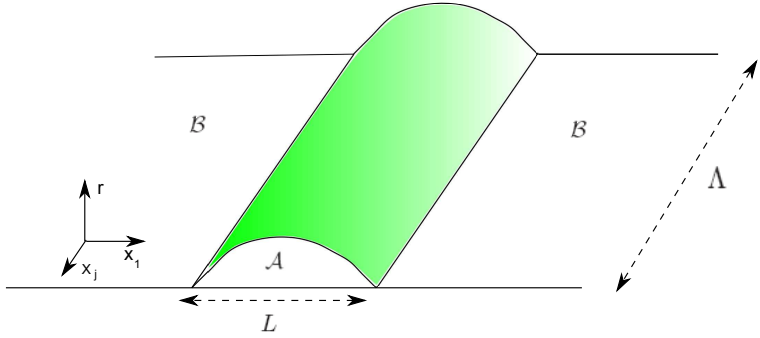}
\caption{Diagram of the stripe shape for the region ${\cal A}$ used to compute the entanglement entropy.}
\label{EEpicture}
\end{center}
%\end{minipage}
\end{figure}

An holographic prescription to compute entanglement entropy (EE) on the AdS$_{d+1}$ gravity dual of a CFT$_d$ was given in \cite{Ryu} in terms of minimal surfaces. The entanglement entropy AdS prescription involves subdividing the system into two regions, ${\cal A}$ and it's complement
${\cal B}$, and find the minimal static $d-1$ dimensional surface (at constant time)
$\gamma_{\cal A}$ such that its boundary coincides with the boundary of the subsystem ${\cal A}$ (see figure \ref{EEpicture}).

The entanglement entropy between the two regions is proposed to be the classical area of $\gamma_{\cal A}$,
\be
{\cal S}_{\cal A}=\frac{2\pi}{\kappa_{(d+1)}^2}\int_{\gamma_{\cal A}} d^{(d-1)}\sigma\sqrt{g^{(d-1)}_{ind}},\label{HEE}
\ee
where $g^{(d-1)}_{ind}$ is the induced metric on the surface and $\kappa_{(d+1)}^2$ is the gravitational constant in $d+1$ dimensions.

In \cite{KlebanovEE} and \cite{PZ}
the EE was computed respectively for backgrounds dual to confining large N gauge theories and for several black holes geometries. In this section we perform this computation for a general background and apply it to the $p$ and $p+ip$-wave superconductors found on section \ref{holsup}. Note that the prescription
to deal with the EE is very similar to that made in \cite{Rey,Malda} to compute VEV of Wilson loops. In the last case the object being
computed is the minimal area of a string that explores the AdS space with its endpoints fixed to the boundary. The following
discussion follows closely that made in \cite{Arias}.

We write the $d+1$ background metric as
\be
ds_{d+1}^2=-g_{tt}(r)dt^2+ g_{x_i x_i}(r)dx_i^2+g_{rr}(r)dr^2,~~~~~~ i=1\ldots d-1,
\ee
where $r$ is the holographic coordinate. The region of interest consists in the straight belt in the direction $x_j$ with width $L$ on the $x_1$ direction. The static embedding belt ansatz is
 $x_1=x_1(\zeta),\, x_j=\zeta_j, r=r(\zeta)$, with $j=2,\ldots, d-1$. A diffeomorphism invariance in ${\cal S}_{\cal A}$ remains, depending on the context it will be fixed either as $x_1=\zeta$ (global embedding) or $r=\zeta$.
The entropy \eqref{HEE} is \be {\cal S}_{\cal
A}=\frac{2\pi\Lambda}{\kappa_{(d+1)}^2}\int d\zeta
\sqrt{g_{x_2x_2}(r)\ldots g_{x_{d-1}
x_{d-1}}(r)}\sqrt{g_{rr}(r)r'^2+g_{x_1x_1}(r)x'^2_1},\label{SA}
\ee where $\Lambda=\int d\zeta_2\ldots d\zeta_{d-1}$ and $'$
denotes $\chi$ derivatives. Defining
$g_{\chi\chi}(r)=g_{x_2x_2}(r)\ldots g_{x_{d-1} x_{d-1}}(r)$ and
the functions \be f^2(r)=g_{\chi\chi}(r)g_{x_1x_1}(r),~~~~~~
\eta^2(r)=g_{\chi\chi}(r)g_{rr}(r) \ee the entanglement entropy is
written \be {\cal S}_{\cal
A}=\frac{2\pi\Lambda}{\kappa_{(d+1)}^2}\int d\zeta
\sqrt{\eta^2(r)r'^2+f^2(r)x'^2_1}.\label{onshellEE} \ee By
minimization of \eqref{onshellEE} we obtain \be x'_1(\zeta)=\pm
\frac{f(r_0)\eta(r)}{f(r)}\frac{r'(\zeta)}{\sqrt{f^2(r)-f^2(r_0)}}\label{rprima},
\ee where $r=r_0$ is the minimum value in the holographic
coordinate reached by the surface. Depending on the background
under study this could be the horizon radius or the end of the
space-time. Inverting this relation we can read the length of the
belt in the x$_1$ direction \be L=2\int_{r_0}^\infty dr
\frac{dx_1}{dr}=2\int_{r_0}^\infty dr
\frac{\eta(r)}{f(r)}\frac{f(r_0)}{\sqrt{f^2(r)-f^2(r_0)}}.\label{Lenght}
\ee We now fix the remaining diffeomorphism invariance as
$x_1(\zeta)=\zeta$, this choice has the advantage
 of providing a complete parametrization of the embedding $r(x_1)$, ($x_1\in[-\frac{L}{2},\frac{L}{2}]$ and the
  boundary condition are $r(\pm\frac{L}{2})=\infty$). Using \eqref{rprima} in \eqref{onshellEE} the entanglement entropy reads
\be
{\cal S}_{\cal A}(r_0)=2\frac{2\pi\Lambda}{\kappa_{(d+1)}^2}\int_{r_0}^\infty dr \frac{f(r)\eta(r)}{\sqrt{f(r)^2-f(r_0)^2}}.\label{connectedsurface}
\ee
Expression \eqref{connectedsurface} diverges at $r=\infty$ due to the infinite extension of the surface. The interpretation of this divergence is that
 another solution exists, with the same boundary conditions, consisting on two
disconnected surfaces expanding all along the radial direction. Its area is
\be
{\cal S}_{{\cal A}_{disc}}=2\frac{2\pi\Lambda}{\kappa_{(d+1)}^2}\int_{r_{min}}^\infty dr\,\, \eta(r),\label{disconnected}
\ee
here $r_{min}$ is the minimum value of $r$ allowed for the geometry. The EE is defined therefore with respect to the reference state \eqref{disconnected}
\be
\Delta {\cal S}_{{\cal A}}=\frac{4\pi\Lambda}{\kappa_{(d+1)}^2}\left(\int_{r_0}^\infty dr \frac{f(r)\eta(r)}{\sqrt{f(r)^2-f(r_0)^2}}-\int_{r_{min}}^\infty dr\,\, \eta(r)\right).\label{DeltaS}
\ee
In what follows we are going to study the EE for the solutions of \eqref{eomsp} and \eqref{pip}.

In the $p$-wave case the relevant functions are:
\be
f^2_p(r)=g_{yy}g_{xx}=r^4,~~~~~~\eta^2_p(r)=g_{yy}g_{rr}=\frac{r^2}{h^2 N}
\ee
where the sub-index $p$ reminds that they correspond to the $p$-wave superconductor.
With this, we can compute explicitly the quantity \eqref{DeltaS}
\be
%{\cal S}_{\cal A}(r_0)&=&\frac{\Lambda}{2G_N^{(4)}}\int_{r_0}^\infty dr \frac{r^3}{h\sqrt{N}\sqrt{r^4-r_0^4}},\\
\Delta {\cal S}_{\cal A}=\frac{4\pi\Lambda}{\kappa_{(4)}^2}\left(\int_{r_0}^\infty dr \frac{r^3}{h\sqrt{N}\sqrt{r^4-r_0^4}}-\int_{r_{min}}^\infty dr\frac{r}{h\sqrt{N}}\right)
\ee

\begin{figure}[ht]
\begin{minipage}{7cm}
\begin{center}
\includegraphics[width=7.5cm]{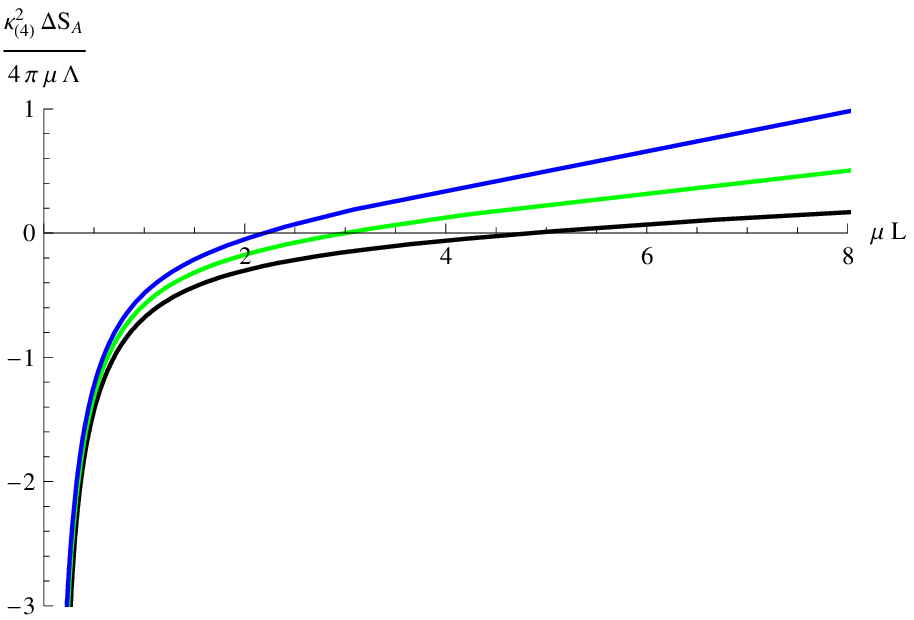}
\caption{ Entanglement Entropy as a function of the size of the strip for the $p$-wave solution. The black, green and blue lines are for values
$g_{_{YM}}=1, T=0.0749\mu$, $g_{_{YM}}=1.5,
T=0.1565\mu$ and $g_{_{YM}}=2, T=0.2312\mu$ respectively.}
\label{EEvsLp}
\end{center}
\end{minipage}
\ \
\hfill\begin{minipage}{7cm}
\begin{center}
\includegraphics[width=7.5cm]{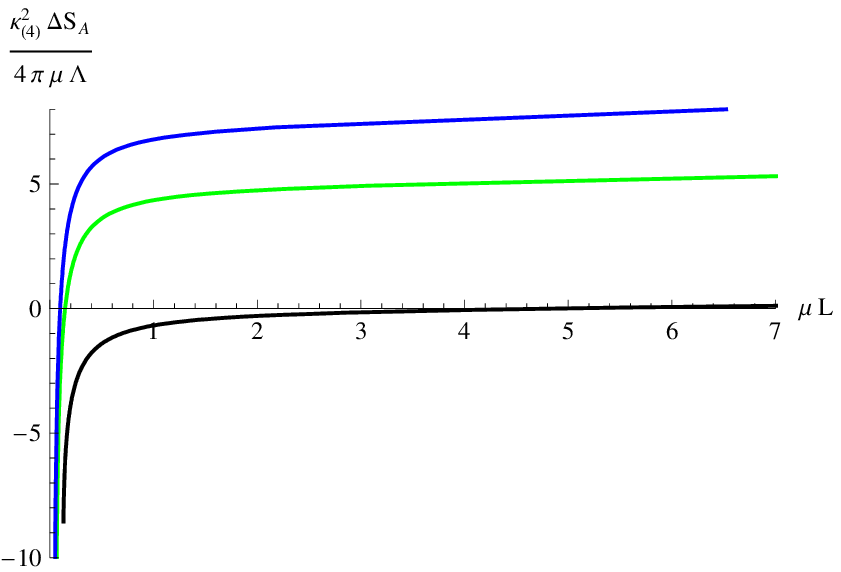}
\caption{Entanglement Entropy as a function of the size of the strip for the $p+ip$ solution. The black, green and blue lines are for values $g_{_{YM}}=1, T=0.0749\mu$, $g_{_{YM}}=1.5,
T=0.1565\mu$ and $g_{_{YM}}=2, T=0.2312\mu$ respectively.}
\label{EEvsLpip}
\end{center}
\end{minipage}
\end{figure}

\begin{figure}[ht]
%\begin{minipage}{7cm}
\begin{center}
\includegraphics[width=7.5cm]{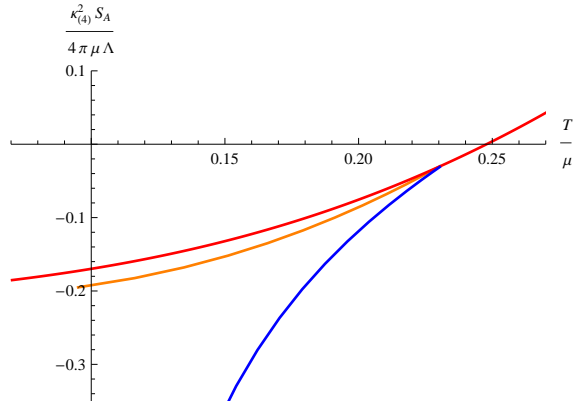}
\caption{ Entanglement Entropy as a function of T and fixed $\mu L=3$, $g_{_{YM}}=2$ and $T_c=0.2312$. The blue line is for the $p$ wave solution, the orange line is for the $p+ip$ superconductor
and the red one is for the Reissner-Nordstrom solution.}
\label{EEvsT}
\end{center}
%\end{minipage}
\end{figure}

On figure \ref{EEvsLp} we plot $\Delta {\cal S}_{\cal A}$ as a function of the length of the strip $L$%, which is possible from invert the equation
%\eqref{Lenght} in order to have $r_0(L)$ and then use this function in \eqref{DeltaS}.
. This shows $\Delta {\cal S}_{\cal A}$ for different values of the
backreaction parameter and different values of the temperature. % The black line is for $g_{_{YM}}=1, T=0.0749\mu$, the green line is for $g_{_{YM}}=1.5, T=0.1565\mu$ and the blue line is for $g_{_{YM}}=2, T=0.2312\mu$.
 As expected the bottom line is the one which has the lowest temperature because as we lower the temperature we must have
more degrees of freedom that condense. The linear behavior for large values of $\mu L$ is a manifestation of the area law proposed in \eqref{HEE}.
On figure \ref{EEvsT} we plot the EE of the condensed (blue line) and normal (red line) phases
as a function of the temperature and for a constant value the length of the belt L. Similar results were found in \cite{Cai2} for the $4+1$ model in
the range of parameters where the second order phase transition arises.

In order to deal with a finite entropy, avoiding the substraction of the disconnected solution, we can write the EE as
\be
{\cal S}_{\cal A}(r_0)=\frac{4\pi\Lambda}{\kappa_{(4)}^2}\int_{r_0}^{\cal R} dr \frac{r^3}{h\sqrt{N}\sqrt{r^4-r_0^4}}=S_{\cal A}+\frac{4\pi\Lambda}{\kappa_{(d+1)}^2}{\cal R}
\ee
where $S_{\cal A}$ has dimensions of length with no divergences. The figure \ref{EEvsT} shows that the EE for the superconductor (blue line)
is lowest that for the RN (red line) solution. This is expected because in the superconducting state there are condensed degrees of freedom.

Performing the same analysis for the solutions of \eqref{pip} we obtain
\be
f^2_{p+ip}(r)=g_{yy}g_{xx}=r^4h^4,~~~~~~\eta^2_{p+ip}(r)=g_{yy}g_{rr}=\frac{r^2h^2}{M},
\ee
and the following EE:
\be
%{\cal S}_{\cal A}(r_0)&=&\frac{\Lambda}{2G_N^{(4)}}\int_{r_0}^\infty dr \frac{r^3h^3}{\sqrt{M}\sqrt{r^4h^4-r_0^4h(r_0)^4}},\\
\Delta {\cal S}_{\cal
A}=\frac{4\pi\Lambda}{\kappa_{(4)}^2}\left(\int_{r_0}^\infty dr
\frac{r^3h^3}{\sqrt{M}\sqrt{r^4h^4-r_0^4h(r_0)^4}}-\int_{r_{min}}^\infty
dr\frac{r h}{\sqrt{M}}\right) \ee In figure \ref{EEvsLpip} we show
the behavior of $\Delta {\cal S}_{\cal A}$ in this case, and we
can perform the same analysis as for the $p$-wave superconductor.
Again, a different approach to obtain a non-divergent entropy, as
in the previous case, avoiding the substraction of the
disconnected surface, consists in separate the divergent piece of
the integral \eqref{connectedsurface} and take in account the
finite part of it, $S_{\cal A}$. In this case: \be {\cal S}_{\cal
A}(r_0)=\frac{4\pi\Lambda}{\kappa_{(4)}^2}\int_{r_0}^{\cal R} dr
\frac{r^3h^3}{\sqrt{M}\sqrt{r^4h^4-r_0^4h(r_0)^4}}=S_{\cal
A}+\frac{4\pi\Lambda}{\kappa_{(4)}^2}{\cal R}. \ee On figure
\ref{EEvsT} we plot this finite part (orange line) for
$g_{_{YM}}=2$ and $T_c=0.2312$ and show that, as expected, is
lower than the EE for the RN solution. Moreover, the figure shows
that the EE in the $p$-wave case is lower than in the $p+ip$
superconductor. This suggest that for a given temperature there
are more condensed degrees of freedom on a $p$-wave
superconductor.

\section{Summary}\label{summary}

On this work we studied the holographic dual to a $p$-wave and a $p+ip$-wave superconductors in 3+1 dimensions. We computed the backreaction of
the $p$-wave solution and studied its thermodynamics properties. As expected, and in contrast with the solution in 4+1 dimensions
studied in \cite{Ammon}, we found a second order phase transition between the normal and superconducting phases. Later, we reviewed the backreaction of the gauge field on the geometry of the colorful black hole and compared it with our solution.
From the study of the thermodynamic quantities and in particular from its grand canonical potential, we noted that for a fixed value of the
temperature the $p$-wave solution has less potential and then it is preferred. We related this with the fact that the $p+ip$ solution is unstable
under small fluctuations and it decays into the $p$-wave background.

Finally, we used the holographic proposal given in \cite{Ryu} to
compute the entanglement entropy of a quantum field theory
studying its gravity dual. We computed it for both solutions on a
straight belt geometry as a function of the temperature and of the
size of the belt. We observed that for both cases the EE behaves
linearly for large values of $L$, which confirms the proposed area
law. The EE vs L plots moves to large values of $\Delta {\cal
S_{A}}$ as we increase the temperature. As a function of the
temperature we observe that the largest EE is for the RN solution.
This is expected because the superconductor has condensed degrees
of freedom. Moreover the $p$-wave solution presents more condensed
degrees of freedom than the colorful black hole. That may explain
the fact that the condensate value is larger for the $p$-wave
system at a given temperature.

\section{Acknowledgments}

We would like to thank Guillermo Silva and Nicolás Grandi for reading the manuscript and to Martin Ammon and Daniel Arean for useful correspondence.
We also thank ICTP, where part of this work was done, for hospitality.
This work has been partially supported by CONICET (PIP2007-0396) and ANPCyT (PICT2007-
0849 and PICT2008-1426) grants. It is also a pleasure to thank FROGS, for inspiration and moral support.

\end{document}